\documentclass[twocolumn,amsmath,amssymb,aps,prl,groupedaddress,floatfix,showpacs,superscriptaddress]{revtex4-1}

\usepackage{graphicx}
\usepackage{dcolumn}
\usepackage{bm}
\usepackage{hyperref}
\hypersetup{colorlinks=true,urlcolor=blue,citecolor=blue}
\usepackage[usenames, dvipsnames]{color} 
\usepackage{braket} 
\usepackage{todonotes}
\usepackage{soul}
\usepackage{tabularx}
\usepackage[caption=false]{subfig}
\usepackage{mathtools}

\begin{document}
\title{Demonstration of a free space rubidium atomic clock with noise below the quantum projection limit} 
\author{Benjamin K. Malia}
\affiliation{Department of Physics, Stanford University, Stanford, California 94305, USA}
\author{Juli\'{a}n Mart\'{i}nez-Rinc\'{o}n}
\affiliation{Department of Physics, Stanford University, Stanford, California 94305, USA}
\author{Yunfan Wu}
\affiliation{Department of Applied Physics, Stanford University, Stanford, California 94305, USA}
\author{Onur Hosten}
\affiliation{Institute of Science and Technology, Austria}
\author{Mark A. Kasevich}
\email[]{kasevich@stanford.edu}
\affiliation{Department of Physics, Stanford University, Stanford, California 94305, USA}
\affiliation{Department of Applied Physics, Stanford University, Stanford, California 94305, USA}
\date{\today}

\begin{abstract}
A technique is demonstrated that allows free space atomic fountain clocks and interferometers to utilize optical cavity generated spin-squeezed states with over $390\,000$ ${}^{87}\text{Rb}$ atoms. Fluorescence imaging is used for population spectroscopy, after a free fall time of 4 milliseconds, to resolve a single-shot phase sensitivity of $814 (61)$ microradians, which is $5.8(0.6)$ decibels (dB) below the quantum projection limit. The dynamic range is observed to be 100 milliradians. When operating as a microwave atomic clock with $240\,000$ atoms at a 3.6 ms Ramsey time, a single-shot fractional frequency stability of $8.4(0.2)\times10^{-12}$ is reported, $3.8(0.2)$ dB below the quantum projection limit.
\end{abstract}

\maketitle

Quantum sensors 
are important for a variety of applications including time keeping and gravitational sensing~\cite{Bloom_2014,Geiger_2011,Rosi_2014}. An atomic fountain clock, for example, works by releasing a cloud of ultra-cold atoms into free space and exposing them to a reference microwave source. A relative phase accumulated due to differences between the microwave and atomic frequency reveals itself in a difference in population between two atomic states~\cite{Wynands_2005}.

Measurements of these population differences in an ensemble of uncorrelated atoms are limited by the quantum projection noise (QPN). For a fixed atom number $N=N_1+N_2$, the variance is such that var$(N_1-N_2)=4$var$(N_1)=4$var$(N_2)\approx N$ if $N_1\approx N_2\approx N/2$~\cite{Wineland_1994}. Surpassing the QPN using quantum correlated initial states of the atoms is an appealing path when increasing the total number of atoms or the interrogation time is technically limited. 

Different approaches exist for preparing entangled states of large atomic ensembles, with spin squeezing --reducing the quantum statistical noise in one of the components of the collective pseudo-spin~\cite{Kitagawa_1993,Wineland_1994}-- being the most successful to date~\cite{Bell2017,QuantumMetrologyReview_2016}. Up to 20 dB in metrologically-relevant variance reduction is obtained by performing quantum non-demolition (QND) measurements in cavity-QED systems with up to one-million atoms~\cite{Hosten_2016,Cox_2016,QuantumMetrologyReview_2016}. These measurements probe the cavity resonance frequency that is modified by the dispersive atom-light interactions, with a sensitivity below the QPN, and reduce the variance of the atomic-spin while preserving its mean. 

Many approaches towards highly-sensitive sensors rely on keeping the atoms confined to an optical lattice during the interrogation time \cite{Bloom_2014}. When squeezing is desired for metrological enhancement in such scenarios, the atoms must be prepared and probed using an optical cavity.

For atomic clock performance, in particular, 10.5(3) dB metrological enhancement was recently shown for a Ramsey time of 228 $\mu$s using 100\,000 atoms~\cite{Hosten_2016}. In-lattice configurations, however, do not allow for long falls in free space since the atoms must be recaptured, limiting the capabilities as inertial sensors. Their performance is degraded due to atom loss and to the spatial-dependent coupling of the expanded atom cloud to the cavity light mode~\cite{Wu_2019_arxiv}. These issues degrade the amount of metrologically-relevant squeezing recovered and limit the performance of the sensor to short interrogation periods. Fluorescence population spectroscopy is a good candidate to allow for much longer interrogation times and to discard the necessity of optical cavity and lattice operation after state preparation. While the state-preparation step is necessarily quantum non-destructive, the subsequent population spectroscopy only requires high fidelity atom counting, and can be destructive. Fluorescence detection of spin-squeezing has previously been used in trapped ions~\cite{Bohnet1297}, while absorption imaging has been preferred for Bose Einstein Condensate systems~\cite{Gross_2010}.

In this Letter we report the first measurement of a cavity-generated spin squeezed state with fluorescence imaging. The method is able to resolve spin squeezed states of $390\,000$ ${}^{87}\text{Rb}$ atoms with up to 5.8 dB in metrologically-relevant variance reduction for a fluorescence collection time of 3 ms using a CMOS camera~\footnote{We note here that the measurements are technically-limited. Atoms are prepared with up to 14 dB of metrological squeezing using QND measurements}. Using this method, we demonstrate a free space atomic clock in an ensemble of 240\,000 atoms with single-shot fractional stability 3.8 dB (2.4-fold) below the QPN limit for a Ramsey time of 3.6 ms, and set the stage for spin-squeezed atom interferometry. 

The  collective pseudo-spin operator in an ensemble of $N$ atoms is the sum of the individual ones, i.e. $J_z=\sum_{i=1}^{N}\sigma_{z,i}/2\equiv (N_\uparrow-N_\downarrow)/2$, where $\sigma_{z,i}$ is the third Pauli operator defined by the clock states $\{\ket{\uparrow}_i,\ket{\downarrow}_i\}$ of the $i$-th atom, and $N_{\uparrow,\downarrow}$ are the number of atoms in the respective states. The coherent, or minimum uncertainty, state of an ensemble of independent atoms with the pseudo-spin pointing in the x-direction ($\langle J_x^{css}\rangle=N/2, \langle J_y^{css}\rangle=\langle J_z^{css}\rangle=0$) is given as the tensor product of all individual states, or $\prod_i (\ket{\uparrow}_i+\ket{\downarrow}_i)/\sqrt{2}$ where CSS stands for Coherent Spin State. This state has a variance of $\langle\Delta J_z^{css}\rangle^2 = \langle\Delta J_y^{css}\rangle^2 = N/4$, which is the QPN value.

Squeezing protocols redistribute the variance of $J_y$ and $J_z$ constrained to the relation $(\Delta J_z) (\Delta J_y) \geq |J_x/2|$. Practically, these protocols can cause a reduction in $J_x$. To account both effects, the metrologically-relevant squeezing is characterized by the Wineland squeezing parameter~\cite{Wineland_1994},
\begin{equation}\label{WinelandPar}
\xi^2 = \Bigg(\frac{\Delta J_z}{\Delta J_z^{css}}\,\,\,\frac{N/2}{|J_x|}\Bigg)^2.
\end{equation}
The first ratio in Eq.~(\ref{WinelandPar}) is the standard deviation reduction $\Xi$, while the second is the inverse of coherence C. As an example, $\xi=0.51$ for the 5.8 dB squeezing recovered in this work.

\begin{figure}[tb]
\centering
\includegraphics[width=\linewidth]{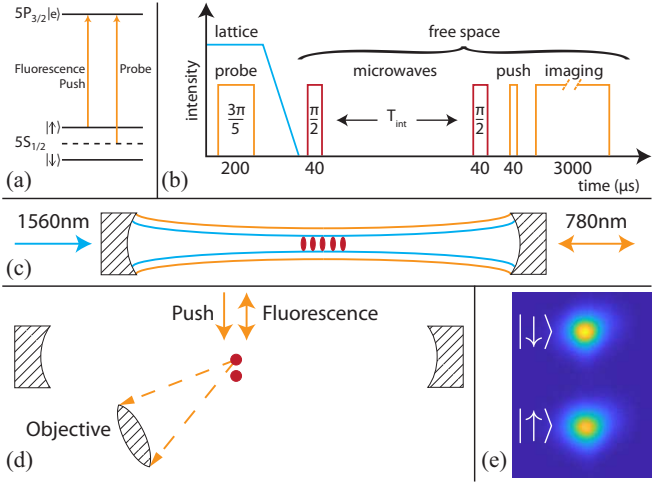}
\caption{Experimental procedures. (a) Relevant atomic states. The probe (QND measurement) light is tuned halfway between the $\ket{\downarrow}$ and $\ket{\uparrow}$ states whereas the fluorescence measurement is tuned directly from the $\ket{\uparrow}$ state to the $\ket{e}$ state. The probe light is on-resonance with the optical cavity. (b) Timing sequence for the experiment. Before the steps shown here, the atoms are prepared in the presqueezed state (see supplement for full sequence). Free fall occurs between the end of the lattice ramp down and the start of the push beam. During clock operation, the two microwave $\pi/2$ pulses are introduced. (c) Atoms are prepared in a horizontal cavity at the potential minima of a one-dimensional 1560 nm lattice. The 780 nm probe squeezes and measures the state. (d) After release, the cloud falls under the force of gravity. The push beam separates the states in space by giving the $\ket{\uparrow}$ state a downward momentum kick. Finally, a vertical, retro-reflected, imaging beam fluoresces the atoms. Repump light (not shown) illuminates the atoms from the side during imaging. (e) A typical image for the separated clouds. The image is 5.3 mm across. The $\ket{\uparrow}$ state is pushed downward in space.}
\label{fig:diagram}
\end{figure}

\begin{figure}[tb]
\centering
\includegraphics[width=\linewidth]{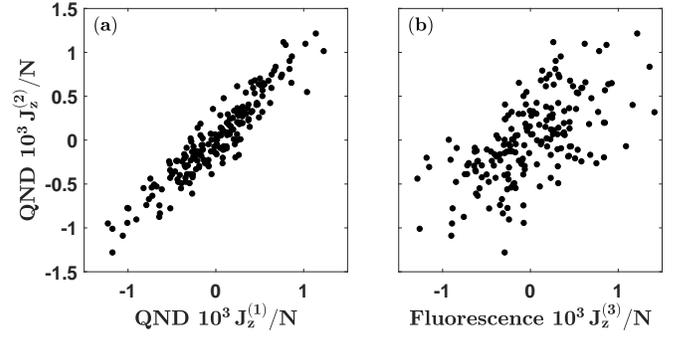}
\caption{Correlation between measurements of $J_z/N$ for $N=390\,000$. Two QND measurements, $J_z^{(1)}$ and $J_z^{(2)}$, and a fluorescence measurement, $J_z^{(3)}$, were taken sequentially. Cavity-cavity measurements (a) show a narrow linear correlation, while cavity-fluorescence measurements (b) show reduced correlations due to additional technical noise associated with fluorescence imaging.} 
\label{fig:correlations}
\end{figure}

\paragraph{}
We generate spin-squeezed atomic ensembles using an almost-homogeneous atom-probe coupling as described by Hosten \textit{et al.}~\cite{Hosten_2016}. The QND measurement has a dual functionality: it prepares the atoms in a 14 dB squeezed state by measuring the $J_z$ component of the ensemble spin and it also serves as a correlation reference to the final fluorescence measurement. A 780-nm beam resonant to the optical cavity is used to probe the atoms (see Fig.~\ref{fig:diagram}a and~\ref{fig:diagram}c). The phase shift on the back reflected light reveals the cavity resonant frequency shift which is measured using a homodyne protocol~\cite{Hosten_2016}. The amount of spin squeezing induced by this measurement is separately calibrated using two correlated QND measurements ~\cite{Schleier-Smith_2010a}.

Fluorescence is used for population spectroscopy after the lattice is turned off and the atoms are released to free space (see Fig.~\ref{fig:diagram}d). A 780-nm laser pulse perpendicular to the cavity state selectively separates the atoms by applying momentum to atoms in the $\ket{\uparrow}$ state. Once the states have separated in space, fluorescence imaging is used to simultaneously count the number of atoms in each state and therefore estimate the $J_z$ component of their collective spin \footnote{This method is similar to the one used in~\cite{Magnification}. However we have improved the photon collection efficiency using a new imaging setup allowing us to perform sub-QPN measurements.}. The fluorescence detection (second measurement) returns a value of $J_z$ within the uncertainty of the non-destructive probe (first measurement). In fact, for a given run $i$, the mean of the second measurement of $J_z$ is set by the output of the first measurement, i.e. $J_{z,2}^{i}=J_{z,1}^{i}+\delta_i$, with $\langle \delta\rangle=0$. This correlation defines the observable of interest to be the difference of the two consecutive measurements, $J_z^{(1,2)} = J_z^{(2)}-J_z^{(1)}$~\cite{Schleier-Smith_2010a}. Although the fluorescence population spectroscopy procedure is noisier than the QND probing, correlations between the two are preserved (see Fig.~\ref{fig:correlations}).

A summary of a typical experimental cycle detailed in ~\cite{Hosten_2016} is outlined here. We generate a $\sim$25 $\mu$K gas of ${}^{87}\text{Rb}$ in an 1560-nm optical dipole lattice trap (Fig. \ref{fig:diagram}c). Atoms distributed in about 1000 sites of a 1D optical lattice are prepared in a coherent 50-50 superposition of the two clock states $\ket{\uparrow} = 5^2S_{1/2}\ket{F = 2, m_F=0}$ and $\ket{\downarrow} = 5^2S_{1/2}\ket{F = 1, m_F=0}$ using a composite-$\pi/2$ low-noise microwave pulse (the microwave system and its phase noise performance is described in the Supplemental Material). The state is then pre-squeezed (one-axis twisting~\cite{Schleier-Smith_2010b}) consecutive with a 200 $\mu$s-long cavity measurement to produce the final squeezing. This probe induces a $3\pi/5$ phase shift on the atomic spins due to the AC Stark effect~\cite{Hosten_2016}. Next, the lattice power is decreased over 200 $\mu$s and the atoms free fall for 4 ms before performing population spectroscopy. At this time, a laser pulse on resonance with the $\ket{\uparrow} \rightarrow  \ket{F'=3}$ transition is applied to the cloud from above for 37 $\mu$s to give a momentum transfer to the atoms in the $\ket{\uparrow}$ state. Once sufficiently separated, a repump beam continuously pumps $\ket{F = 1} \rightarrow \ket{F = 2}$ through $\ket{e}$. A simultaneous retro-reflected imaging beam 0.2 $\Gamma$ red detuned from the $\ket{F = 2} \rightarrow \ket{e}$ transition induces $\ket{e} \rightarrow \ket{F = 2}$ emission in both clouds for 3 ms. Here $\Gamma$ is the atomic linewidth, $(2\pi)\times6.06$ MHz. 

The fluorescence imaging system consists of an objective lens with a numerical aperture of 0.25 and a 1:1 relay to a camera. The sensor is a commercially available low read noise CMOS camera (supplement). Each cloud is spatially well resolved on the 3.1 Mpx camera without overlap between them. Typical cloud sizes are 3 mm in diameter and remained independent of atom number. These imaging parameters are limited by the field of view of the camera and cloud expansion over time. Atom number is calibrated to the probe measurement by counting electrons generated in the camera for a given cavity shift. The camera detects on average 65 photons per atom over the 3 ms exposure time. We correct for a slight pixel-dependence on the collection efficiency (described in supplement).

Rabi oscillation sequences for the CSS atomic ensemble reveal a coherence of $C \approx 98\%$~\cite{Hosten_2016}, squeezed-state preparation with the QND probe reduces this to $C \approx 91\%$ (see Tab.~\ref{tab:coherence} and supplemental material). The coherence remains at this level as the atoms are released from the lattice. In the described configuration, the total free fall time – between release and spatial state separation – available to utilize the atomic spins for metrology is 4 ms due to atoms leaving the field of view of the camera. However in an alternative configuration by increasing the time over which the lattice power decreases to 7ms, the cloud can be made to expand more slowly and the camera can collect images after free fall times up to 8ms. These long free falls are possible at the cost of coherence ($C\approx 73\%$) caused by the longer exposure to the lattice potential, which results in less metrologically-relevant squeezing. This is our main limitation to achieving interrogation times longer than 8 ms. Remarkably, atoms preserve their coherence as well as the squeezing levels for all free-fall times explored in this work. This result shows that any disturbance of the squeezed states by the environment during free fall is below the detection limit of the camera. For the purpose of achieving the largest squeezing at long times, all of the data presented below were taken using a lattice time of 0.2 ms and a free fall time of 4 ms.

\begingroup
\squeezetable
\begin{table}[t]
\begin{ruledtabular}
\begin{tabularx}{\linewidth}{@{\extracolsep{\fill} }lrrr@{\hskip .5cm}rrrr}
Lattice time (ms) & \multicolumn{3}{c}{0.2} & \multicolumn{4}{c}{7.0}\\

Free fall time  (ms) & 2& 3& 4& 2& 4& 6& 8\\

C ($\pm$0.3)(\%)& 91.0 & 90.6 & 91.7 & 72.7 & 73.5 & 73.6 & 73.8\\

$\Xi^2$ ($\pm$0.6)(dB)& -5.9 & -7.0 & -6.6 & -6.0 & -6.4 & -7.2 & -6.9\\

$\xi^2$ ($\pm$0.6)(dB)& -5.1 & -6.2 & -5.8 & -3.2 & -3.7 & -4.5 & -4.3\\
\end{tabularx}
\end{ruledtabular}
\caption{State coherence $C=|J_x|/(N/2)$, variance reduction $\Xi^2=\textnormal{var}(J_z)/(N/4)$, and metrologically-relevant squeezing $\xi^2=\Xi^2/C^2$, as a function of free fall time for both short and adiabatic release times. $N=390\,000$ atoms.}
\label{tab:coherence}
\end{table}
\endgroup

\begin{figure}[tb]
\centering
\includegraphics[width=\linewidth]{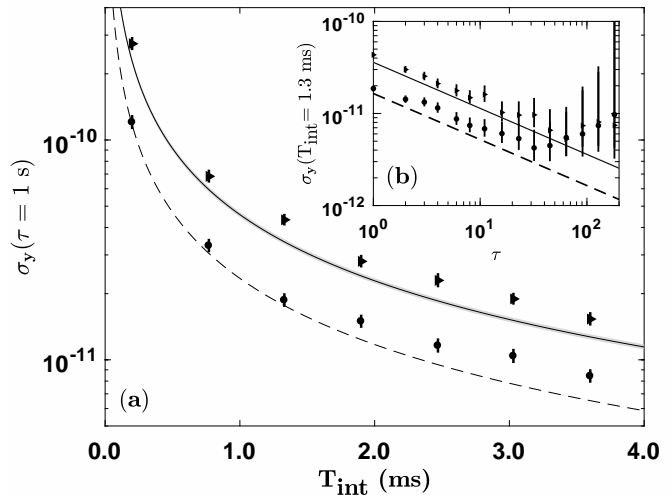}
\caption{Circles: squeezed state. Triangles: CSS. All error bars represent 99\% confidence intervals for visual clarity. (a) Single-shot fractional stability $\sigma_y(\tau=T_c=1\text{ s})$ for several Ramsey times $T_{\text{int}}$ in a clock sequence with 240\,000 atoms. Solid line: QPN limit, eq.~(\ref{eq:stability}). Dashed line: 5.8 dB squeezed clock for reference. (b) Fractional stability for $T_{\text{int}}$ = 1.3 ms as a function of averaging time $\tau$. }
\label{fig:allanDev}
\end{figure}

\begin{figure}[tb]
\centering
\includegraphics[width=\linewidth]{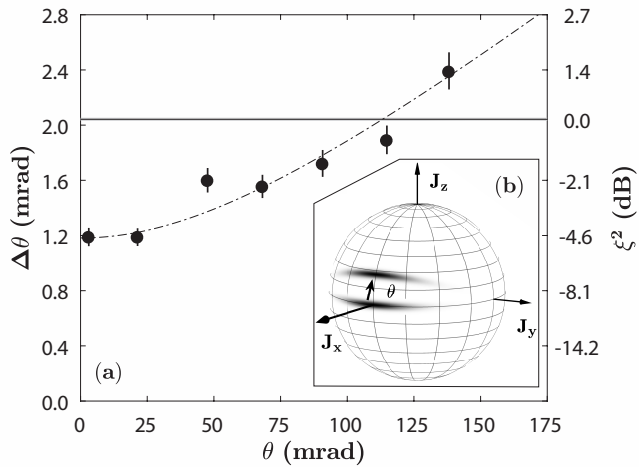}
\caption{{}(a) Angular sensitivity vs induced rotation angle for 240\,000 atoms. Error bars represent 68\% confidence intervals. Solid line: QPN limit, $1/\sqrt{N}$. Dashed line: theoretical fit to a squeezed state with excess anti-squeezing rotated on the Bloch sphere (see suplemental material for details). Right size scale shows metrologically-relevant squeezing. (b) Bloch sphere representation of a rotated squeezed state. Figure is not to scale; the true aspect ratio of the squeezed states in the experiment is approximately 300:1.}
\label{fig:angularResolution}
\end{figure}

\paragraph{}
We evaluate the performance of the system to resolve small tipping angles off the equator on the spin-1/2 multi-particle Bloch sphere (Fig.~\ref{fig:angularResolution}). Measurements of such small angles are crucial for quantum state discrimination techniques and adaptive protocols for increasing atomic clock performance~\cite{Adaptive}. These angles are along the polar direction. and are estimated by dividing the spin measurements by the effective radius of the sphere, $\theta = J_z^{(1,2)}/C(N/2)$. All reported squeezing values are relative to a CSS with full contrast ($C=1$). 

For states near the equator, the best resolution, $\Delta\theta$, of the system is $814(61)$ $\mu$rad when using 390\,000 atoms for a free fall time of 4 ms. These values correspond to a maximum of $5.8(0.6)$ dB squeezing. Even though we are able to prepare 14 dB squeezed states, the fluorescence resolved values are limited to 5.8 dB due to 740 $\mu$rad of technical noise. Some contributions to this noise come from background scattered light, and photon shot noise. A detailed description of noise sources is provided in the supplement.

We now utilize the squeezed states in a standard clock sequence. $240\,000$ atoms are released from the lattice and rotated to a phase sensitive state by a $\pi/2$ microwave pulse about the $J_x$ axis (Fig.~\ref{fig:diagram}b). After some time $T_{\text{int}}$, the atoms accumulate a phase relative to the microwave source and are then rotated back with another $\pi/2$ pulse. This pulse is phase shifted by roughly 180$^\circ$. This procedure maps accumulated phase to population difference and the atoms are later state separated and imaged. In the case of a clock with CSSs, the atoms are prepared in the $\ket{\downarrow}$ state and two $\pi/2$ microwave pulses separated by a $\pi/2$ phase shift are applied.

The performance of an atomic clock is characterized by the uncertainty of the accumulated phase, or frequency fractional stability $\sigma_y$. Assuming full contrast, the QPN limit to the stability takes the form~\cite{Wynands_2005}
\begin{equation}\label{eq:stability}
\sigma_y^{(css)} = \frac{1}{\omega_0 T_{\text{int}}}\sqrt{\frac{T_c}{N\tau}},
\end{equation}
where $\omega_0=2\pi\times6.834$ GHz is the atomic transition frequency, $\tau$ is the averaging time, $T_{\text{int}}$ is the Ramsey or interrogation time, and $T_c$ is the experimental cycle time.

For a Ramsey time of 3.6 ms, for example, we observe a single-shot fractional frequency stability ($\tau=T_c=1$ s) of $8.4(0.2)\times10^{-12}$ (Fig. \ref{fig:allanDev}a). This corresponds to a $3.8(0.2)$ dB metrological improvement in clock performance below the QPN limit, or a 2.4 times reduction in averaging time. For $T_{\text{int}}\leq1.3$ ms, the clock is limited by the resolution of the fluorescence imaging. Furthermore, we observe that the fractional stability can, in general, be averaged down while remaining below the QPN limit (Fig. \ref{fig:allanDev}b). Microwave amplitude and phase noise as well as magnetic field fluctuations limit the ultimate resolution of our system to $\sigma_y\approx4\times10^{-12}$.

The resolution of these types of sensors can be hindered by the inherent increase in $\Delta J_y$, or anti-squeezing. Anti-squeezing makes $\Delta J_z$ sensitive to rotations on the Bloch sphere~\cite{Braverman_2018}. At large angles above the equator, the state's downward curvature projects onto $J_z$ (Fig. ~\ref{fig:angularResolution}b). This issue can be tackled using adaptive measurements to rotate the state close to the equator before a final measurement is performed~\cite{Adaptive}. Here we evaluate the dynamic range of our sensor; outside of which, adaptive methods would be required. We perform the above clock sequence for $240\,000$ atoms and $T_{\text{int}} = 10~\mu s$ with an additional phase shift of $\theta$ between the two pulses. This sequence places the state at an angle $\theta$ above the equator. The states are anti-squeezed by 36 dB due to low quantum detection efficiency of the homodyne setup~\cite{Hosten_2016}, and the clock sequence adds an extra 590 $\mu$rad of technical noise. Figure~\ref{fig:angularResolution}a shows the dynamic range of the measurement. Metrologicaly-relevant squeezing can be utilized for rotations up to $\pm125$ mrad, exceeding by almost two orders of magnitude the dynamic range of the QND measurement, which is limited to angles of $\pm1.6$ mrad~\cite{Wu_2019_arxiv}. 

Our results demonstrate that the fluorescence measurement method is capable of resolving highly squeezed cold atom spin states. This work paves the way for incorporating highly squeezed states into an atom interferometer. For example, a cold-atom gravimeter detects phase shifts of $\theta\propto gT^2$ where $g$ is the local acceleration and $T$ is the free fall time~\cite{Rosi_2014}. Our method would be approximately twice as sensitive as a QPN-limited sensor for the same $T$. To take advantage of long $T$, atoms would need to be prepared in a colder configuration to reduce the expansion of the cloud. 

\begin{acknowledgments}
Work supported by the Office of Naval Research, Vannevar Bush
Faculty Fellowship Department of Energy, and Defense Threat Reduction Agency.
\end{acknowledgments}

\clearpage

\renewcommand{\subsectionautorefname}{\S}
\renewcommand{\sectionautorefname}{\S}

\renewcommand\theequation{S\arabic{equation}}
\setcounter{equation}{0}
\renewcommand\thefigure{S\arabic{figure}}
\setcounter{figure}{0}
\renewcommand\thetable{S\arabic{table}}
\setcounter{table}{0}
\renewcommand\thesection{S\arabic{section}}
\setcounter{section}{0}
\renewcommand\thesubsection{\thesection.\arabic{subsection}}
\setcounter{subsection}{0}
\setcounter{secnumdepth}{2} 
\bibpunct{[S}{]}{, S\!\!}{n}{}{,}
\makeatletter
\renewcommand\@biblabel[1]{[S#1]}
\makeatother

\section*{SUPPLEMENTARY MATERIAL}
\section{State Preparation}
\begin{figure}[ht]
\centering
\includegraphics[width=0.8\linewidth]{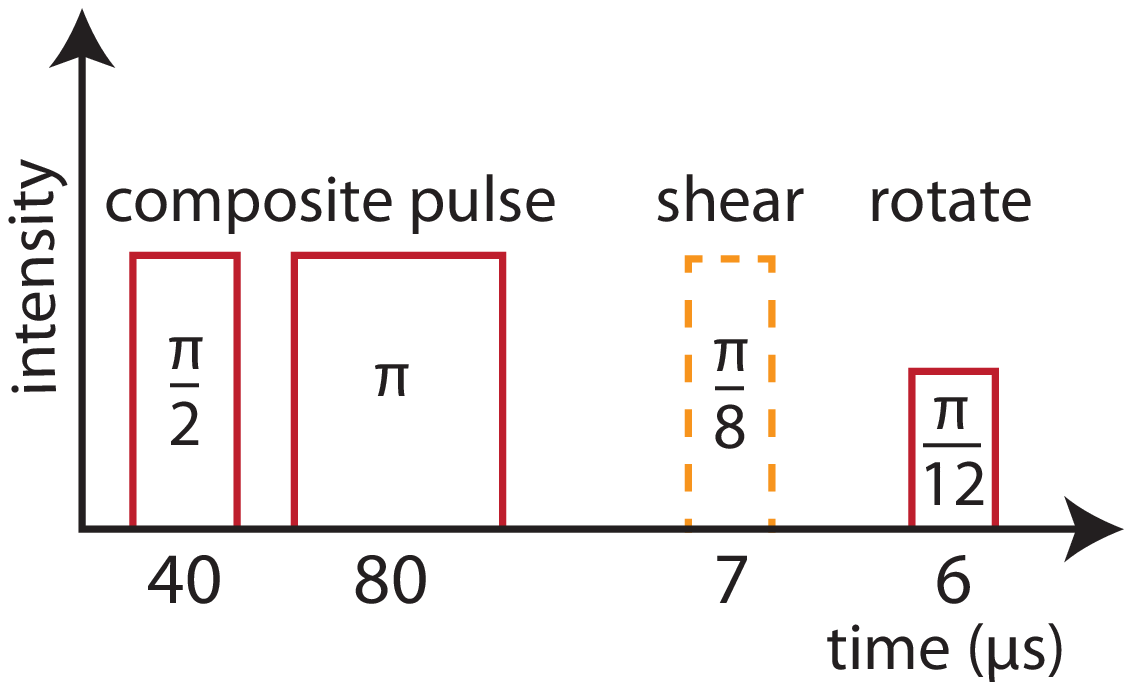}
\caption{First, a composite $\pi/2$ microwave pulse prepares the $J_z=0$ state. The detuned probe light shears the state and a final microwave pulse rotates the state to align with the equator.}
\label{fig:stateprep}
\end{figure}

\begin{figure}[ht]
        \includegraphics[width=0.8\linewidth]{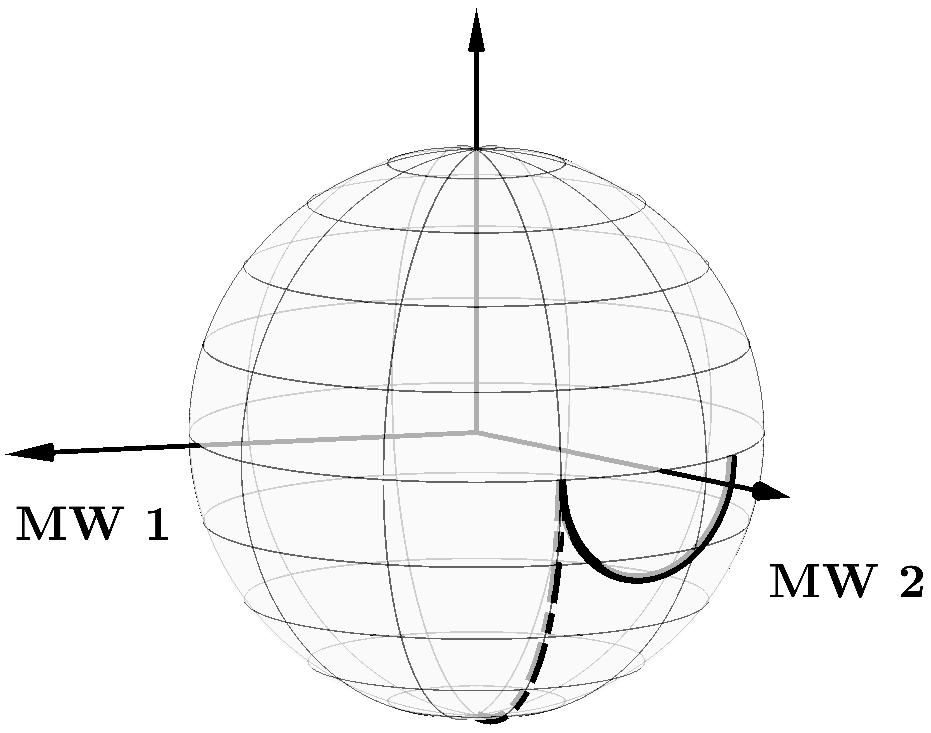}
    \caption{Composite microwave $\pi/2$ pulse for preparing the state at the equator ($J_z = 0$ state). MW1 $\pi/2$ pulse (dashed line) brings state from south pole ($\ket{\downarrow}$) near the equator. MW2 $\pi$ pulse (solid line) brings the state closer to the equator. The two rotation axis are $120^\circ$ degrees apart.}
    \label{fig:composite}
\end{figure}

The state preparation sequence is shown in Fig.~\ref{fig:stateprep}. To prepare a coherent state, atoms are first placed in the ground state and all remaining atoms in the $\ket{\uparrow}$ state are pushed away so that only $\ket{\downarrow}$ remain. Then to rotate this state to the equator of the Bloch sphere, a composite scheme was used to reduce amplitude noise from the microwave source (Fig.~\ref{fig:composite}). In this composite sequence, a $\pi/2$ pulse brings the $\ket{\downarrow}$ state to the equator, then a $\pi$ pulse $120^\circ$ phase-shifted from the first rotates the state back to the equator on the other side of the Bloch sphere. If, for example, in a given run the microwave power was low, the state would be below the equator after the first $\pi/2$ -weak- pulse. The second $\pi$ pulse would be weak as well, and put the state on the equator, instead of passing it. The pairs of pulses used to determine the fractional stability $\sigma_y$ and $\Delta \theta$ for $\theta>0$ act in a similar way to reduce the amplitude noise of the microwave system.

\paragraph{}
Once the coherent state is at the equator ($J_z=0$), we perform pre-squeezing. This procedure utilizes one-axis twisting to generate non-conditional squeezing~\cite{Hosten_2016}. A weak $\pi/8$ probe $6.25\kappa_0 \approx 50$ kHz detuned from the cavity resonant frequency shears the state through the Hamiltonian $H \propto J_z^2$. This interaction is followed by a $\pi/12$ microwave rotation about the axis of the atoms to align the elongated state with the equator of the Bloch sphere. The posterior QND measurement reveals that these states are squeezed 6 dB below the QPN limit. This procedure allows us to prepare large ensembles of atoms within the linear response of the cavity and to collect data more efficiently.

\section{Simultaneous Imaging}
Imaging both populations at the same time offers several advantages~\cite{Rocco_2014}. Imaging light intensity and frequency can vary shot-to-shot, thereby changing the number of photons fluoresced per atom. Since both clouds are imaged simultaneously, these fluctuations look like total atom number fluctuations. Any variation in the gain of the sensor would also have the same effect. The total measured atom number fluctuates by about $\pm 5\%$, being the largest fluctuation in the experiment. Reported atom numbers are the mean atom number of a data set.

To eliminate these variations as well as true total atom number fluctuations, we compute $J_z$ in each data run using the normalized expression 
\begin{equation}
J'_z = \frac{\bar{N}}{2}\frac{(N_\uparrow - N_\downarrow)}{(N_\uparrow + N_\downarrow)},
\end{equation} 
where $N_i$ is atom number in state $i$ and $\bar{N}=\bar{N_\uparrow}+\bar{N_\downarrow}$ is the mean atom number of the whole data set. This formula eliminates $\Delta N$ contributions to $\Delta J_z$. For example, for a coherent state such that $\text{var}(N_\uparrow)=\text{var}(N_\downarrow)=\bar{N}p(1-p)$, the variance takes the form  
\begin{eqnarray}
    \text{var}(J'_z) &=& \bar{N}p(1-p)+(1-p)^2 \, \text{var}(x_\downarrow) +p^2 \,\text{var}(x_\uparrow)\nonumber\\&&-2p(1-p) \,\text{cov}(x_\downarrow,x_\uparrow),
\end{eqnarray}
where $p=\bar{N_\uparrow}/\bar{N}$ such that $\bar{J'_z}=(1-2p)\bar{N}/2$, and $x_i$ is the amount of background counts due to scattered light for state $\ket{i}$. We note that  $\text{var}(x_\uparrow)\approx \text{var}(x_\downarrow)$. The first term is the noise of the coherent state, the middle two terms are the noise from background scattered light, and the last term is the correlation of the background scattered light due to simultaneous imaging. 
Noise from the background light contributes about -11 dB to the measurement of 240,000 atoms at $\theta = 125$ mrad. Since the measured noise is near the QPN at this angle, use of the normalized formula is not a limitation on this method. As a comparison we give the uncertainty for a non-normalized expression, showing the dependence on $\text{var}(N)$
\begin{eqnarray}
    \text{var}\left(\frac{N_\uparrow-N_\downarrow}{2}\right) &=& \bar{N}p(1-p)+\left[\frac{1}{4}-p(1-p)\right]\text{var}(N)\nonumber\\
    &+& \frac{\text{var}(x_\downarrow)}{4}+\frac{\text{var}(x_\uparrow)}{4} -\frac{\text{cov}(x_\downarrow,x_\uparrow)}{2}.\nonumber
\end{eqnarray}

\section{Camera system construction}
\paragraph{}
The FLIR Blackfly S model BFS-U3-32S4M-C contains a 7.0mm by 5.3mm Sony IMX252 CMOS sensor with 3.45$\mu$m size pixels (3.2Mpx total). The camera is attached to the end of a 1"-diameter tube containing two $f=50$ mm lenses placed $2f$ apart. This relay creates a 1:1 magnification insensitive to slight position misalignment of the camera~\cite{Pyragius_2012}. A Semrock 780 nm laser clean-up filter is placed immediately after the objective lens. The objective lens rests approximately $f$ from the atoms and is about 1" in diameter (NA $\approx 0.25$).
\paragraph{}
Our collection efficiency was limited by the quantum efficiency of our camera (35\%) as well as our relatively large distance (5cm) of the 1" objective lens to the atoms. The imaging could be improved by using a cmos camera and imaging system with an increased field of view and a higher photon collection efficiency since both background light and read noise scale with the collection efficiency (see section ~\ref{sec:noise}. Furthermore, implementing colder atoms would reduce cloud expansion limitations.

\section{Background subtraction and characterization}
\paragraph{}
Background light consists mostly of imaging and repump beams scattering off of the vacuum chamber zerodur walls. To reduce the number of scattered photons, only a vertical imaging beam and a horizontal repump beam were used. Both beams are transverse to the field of view of the camera.
\paragraph{}
After the image of the atoms is taken, the imaging and repump light is turned off and the atoms fall out of the field of view. The light is then turned back on and a second image is taken 16ms after the first. This is the shortest possible time using the camera's trigger mode. The two images are subtracted to eliminate mean offset of the background light. The mean background counts after subtraction account for less than 0.1\% of the total counts in the image. 

\section{Camera noise sources}
\label{sec:noise}
\paragraph{}
Noise sources other than quantum uncertainty of the squeezed atomic state come from multiple sources and are summarized in \autoref{tab:noise}.

\begingroup
\squeezetable
\begin{table}[t]
\begin{tabular}{lc}
\hline
\hline
Source & Noise \\
&(dB)\\
\hline

unidentified & -11 \\
read noise \& background scattered light & -14\\
trap inhomogeneity & -16\\
photon shot noise & -18\\
\hline
total & -8\\
\hline
\hline
\end{tabular}
\caption{Approximate noise contributions to $\Delta \theta$, relative to the QPN, for N = 390\,000. The fluorescence measured variance reduction $\Xi^2 =-6.6(0.6)$ dB is a combination of the true quantum state noise ($\approx -14$ dB) and $\approx -8$ dB of additional noise sources.}
\label{tab:noise}
\end{table}
\endgroup

\paragraph{}
Photon shot noise arises from the total number of photons $N_\gamma = \alpha N$ collected from each state, where $\alpha$ is the average number of photons collected per atom. The photon noise contribution is therefore
\begin{equation}
    \Delta J_z^\gamma = \frac{1}{\sqrt{\alpha}}\frac{\sqrt{N}}{2},
\end{equation}
giving a noise contribution of $20\log(1/\sqrt{\alpha})\approx -18$ dB.

\paragraph{}
Background counts in the image arise due to camera read noise and background scattered light. The readnoise is much smaller than the scattered light. Background noise contribution is

\begin{equation}
    \Delta J_z^\text{background} = \frac{\Delta X}{\alpha}
\end{equation}
where $X = (x_\uparrow - x_\downarrow)/2$. These sources are suppressed by increasing photon emission rate from the atoms while amplified by increasing the number of background scattered photons. The fluorescence imaging power was therefore chosen to minimize $\Delta J_z^\text{background}$. Total contribution from this source is -14 dB for $N = 390\,000$. 

\paragraph{}
While atom-probe coupling in the cavity is near uniform during state preparation, there are still small variations due to the position of the atoms in the trap. Thermal fluctuations can cause these variations during the QND measurement. The additional noise is given by ~\cite{Hosten_2016}:

\begin{equation}
    \Delta J_z^\text{thermal} = \frac{\sqrt{N}\beta^2}{\sqrt{1+2\beta^2}}
\end{equation}
where $\beta = 2\sigma_r/\omega$ with $\sigma_r$ the transverse size of the atomic cloud and $\omega$ the probe beam waist. In this experiment, $\beta^2 = 0.076$. The fluorescence measurement is not sensitive to these positional fluctuations and therefore does not contain this type of noise. Total contribution from this source is $20\log(2\beta^2/\sqrt{1+2\beta^2})\approx-16$ dB.

\section{Photon collection and position corrections to $J_z$}
\begin{figure}[ht]
        \includegraphics[width=\linewidth]{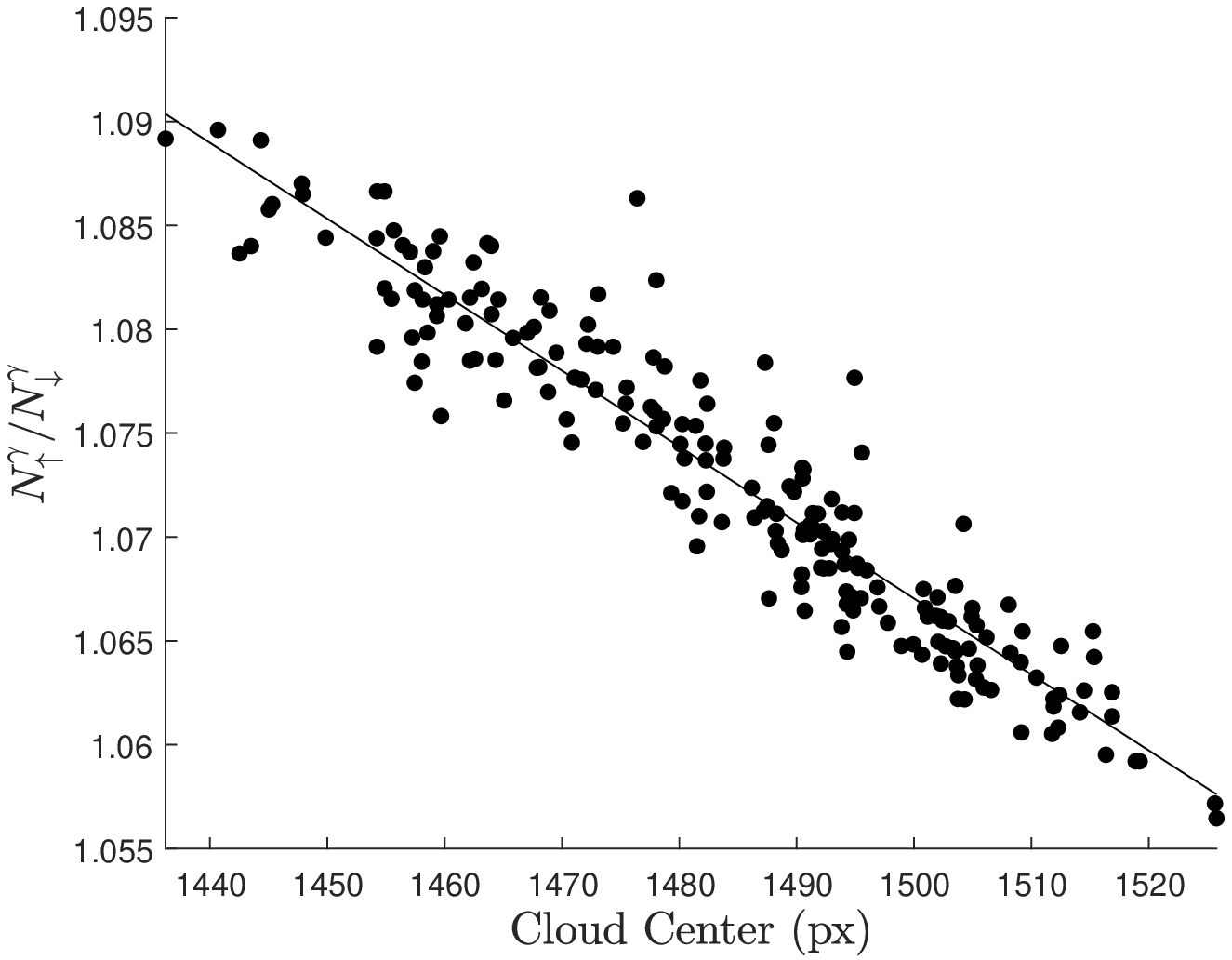}
    \caption{Ratio of photon counts between both states as a function of pushed cloud's center.  Solid line indicates a linear fit to the data.}
    \label{fig:position}
\end{figure}

\paragraph{}
Scattered light from the unpushed cloud was calibrated to the atom number determined from cavity measurements.  This was done by measuring the cavity resonance frequency shift for clouds prepared in the ground state and correlating it with number of detected photons.  Using the specified quantum efficiency of the sensor is (0.35 $e^-$/photon) and the sensor gain we estimate that $\sim$ 65 photons are detected per atom. 
\paragraph{}
Due to shot-to-shot intensity fluctuation of the push beam, the position of the pushed cloud varied by up to $\sim$ 500 $\mu$m shot to shot. This position dependence is correlated with the photon collection efficiency (Fig ~\ref{fig:position}). To correct for this effect, we calibrated the fluorescence measurement ratio of the two clouds with a squeezed state initially prepared on the Bloch sphere equator.  From this data, we determined a correction factor (as a function of position) so that the mean $J_z$ measured by the camera was zero. The correction modifies the photon count in the pushed cloud only.

\section{Beatnote correction}
The Hamiltonian for the interaction of the probe light with the atoms is~\cite{Hosten_2016}
\begin{equation}
    H = \frac{1}{2}g^2a^\dagger a\Bigg(\frac{N_\uparrow}{\Delta_\uparrow}-\frac{N_\downarrow}{\Delta_\downarrow}\Bigg),
\end{equation}
where $a^\dagger$ and $a$ are the photon creation and annihilation operators, $g$ is the coupling constant, and $\Delta_i$ is the detuning between the probe laser and the state $i$. The measured $J_z$ therefore depends on the frequency detuning of the probe.  In our experiment, the 780 nm probe light is frequency locked to the cooling light.  However, the beatnote between the two can fluctuate by up to $\pm 3$ MHz.  We correct for this fluctuation by measuring the beatnote shift $\delta$ and computing the inferred $J_z$ value of the cavity via the equation

\begin{equation}
    J_z^{\text{inferred}} = J_z^{\text{measured}} + \frac{1}{2}\frac{N\delta}{\Delta},
\end{equation}
where $\Delta = 3.417$ GHz is the mean detuning. This equation is valid for small $\delta/\Delta$, where $\Delta_\uparrow=\Delta+\delta$ and $\Delta_\downarrow=\Delta-\delta$.  Note that this effect does not reveal itself when cavity measurements alone are used to produce and then read-out squeezed states, since it is common to both the preparation and read-out processes.

\section{Post-selection}
Approximately 10-15\% of data are removed from the analysis for several reasons. First, states with $J_z$ values outside of the linear response range of the cavity ($\pm 1$ kHz $\Rightarrow J_z = \pm160$) are removed. Since the observable of interest is $J_z^{(1,2)}$, selectively removing points based on the first measurement value only does not change the accuracy of the result. Second, in the clock measurement technical noise creates anomalous results in less than 1\% of the data. These are eliminated by removing $J_z$ values over $\pm6\times$QPN levels away from the mean.

\section{Coherence Measurement}
\begin{figure}[t]
    \includegraphics[width=\linewidth]{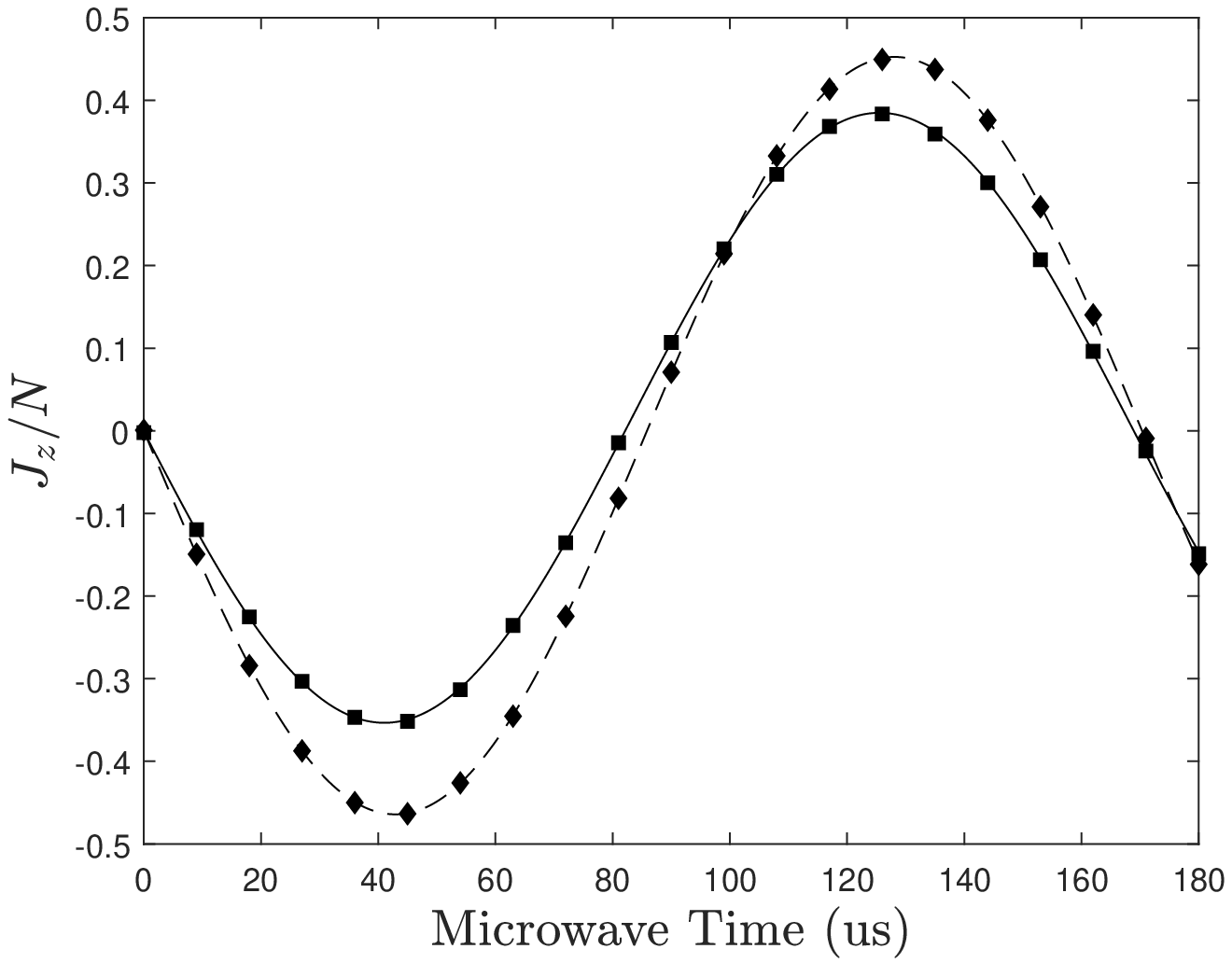}
    \caption{Rabi sequence with 4 ms free fall time and 0.2 ms lattice time (diamonds), and 8 ms free fall time and 7.0 ms lattice time (squares). $C= 0.91 \text{ and } 0.73$ are calculated from the respective fitted curves.}
    \label{fig:coherence}
\end{figure}

Coherence is determined by preparing a squeezed state on the equator and then aligning the MW axis to the atoms. After a $\pi/2$ phase shift, a microwave pulse 200 $\mu$s before the push beam is applied to the atoms. The microwave pulse area is increased until a full oscillation is observed (Fig.~\ref{fig:coherence}). This waveform is used to determine $C$. 

\section{Uncertainty on Noise Measurements}
The noise $\Delta J_z$ is calculated by the standard deviation of the data set. We calculate error bars on these noise measurements via

\begin{equation}
    \sqrt{\frac{(n-1)\Delta J_z^2}{\chi^2_{n-1,1-\frac{m}{2}}}}<\Delta J_z< \sqrt{\frac{(n-1)\Delta J_z^2}{\chi^2_{n-1,\frac{m}{2}}}},
\end{equation}
where $\chi^2_{d,\sigma}$ is value of the $\chi^2$ distribution with $d$ degrees of freedom at $\sigma$. $m = 0.68$ represents a confidence interval of $\pm1$ standard deviation from the mean ~\cite{Krishnakumar_2017}. $n \approx 200$ is the number of samples in the data set. Each reported result is a combination of four data sets. In the above formula, $n$ is taken as the smallest $n_i$ of the four sets and $\Delta J_z$ as the pooled variance,

\begin{equation}
    \Delta J_z^\text{pooled} = \frac{\sum_i [(n_i-1)(\Delta J_z)_i]}{\sum_j (n_j-1)}.
\end{equation}

\section{Microwave Phase Noise}
We generate the microwave transition frequency between $\ket{F=1,m_F=0}$ and $\ket{F=2,m_F=0}$ states ($\sim6.834$ GHz) by first mixing a $\sim 7.2$ GHz signal from a Wenzel Golden Multiplied Crystal Oscillator (P/N: 501-30459) with a $\sim 366$ MHz signal from a direct digital synthesizer (AD9914). This mixed signal is then filtered by a band-pass filter to give a $\sim 6.834$ GHz component. This $\sim6.834$ GHz component is amplified, and $\sim 1$ W of radio frequency power is sent to the atoms through a microwave horn. The reference clock frequency for this direct digital synthesizer is $\sim2.4$ GHz which is generated from the same Wenzel Oscillator as the $\sim7.2$ GHz signal. This Wenzel Oscillator is locked to a $10$ MHz crystal oscillator (HP10811-60111) with a loop bandwidth $< 10$ Hz. To characterize the phase jitter of this microwave system, we built two identical copies of this system and mixed the signals from the two copies.  From these measurements, we expect an additional contribution of $-10$ dB of noise (relative to the shot-noise level).  This accounts for half of the additional noises we have measured in this experiment.

\section{Allan Deviation}
For evaluating clock performance, the variable of interest is the ratio between measured and average accumulated microwave phase,
\begin{equation}
    \phi(t) = \frac{\theta(t)}{T_{int}\omega_0} = \frac{J_z^{(1,2)}(t)}{C(N/2)T_{int}\omega_0},
\end{equation}
where $\omega_0=(2\pi)\times6.834$ GHz is the transition frequency between the states, $N$ is the mean number of atoms, $C$ is the Rabi coherence, and $T_{int}$ is the Ramsey time. 

The fractional stability of the clock is defined as
\begin{equation}
\sigma_y^2(\tau) = \frac{1}{2(M-1)}\sum_{k=1}^{M-1} (\bar{y}_{k+1} - \bar{y}_k)^2
\label{allan_dev}
\end{equation}
where $\tau$ is the averaging time, $M=\text{floor}(T/\tau)$ is the number of averaged samples, $T$ is the total data collection time, and 
\begin{equation}
\bar{y}_k = \frac{1}{L}\sum_{i=1}^{L} \phi((k-1)\tau+iT_c),
\label{y_bar}
\end{equation}
with $L=\tau/T_c$ the size of each sample and $T_c$ the experimental cycle time.

Any post-selected points $\phi(t)$ are not included in the calculation of $\bar{y}_k$ (\ref{y_bar}). This may result in some $\bar{y}_k$ without any $\phi(t)$. In this case, terms in \ref{allan_dev} which contain these special $\bar{y}_k$ are not included in the calculation of $\sigma_y^2(\tau)$. Expressions for $\bar{y}_k$ and $\sigma^2_y$ are normalized accordingly. This method preserves the time information of the data without eliminating large sections of data.

\begin{figure}[ht]
\centering
\includegraphics[width=\linewidth]{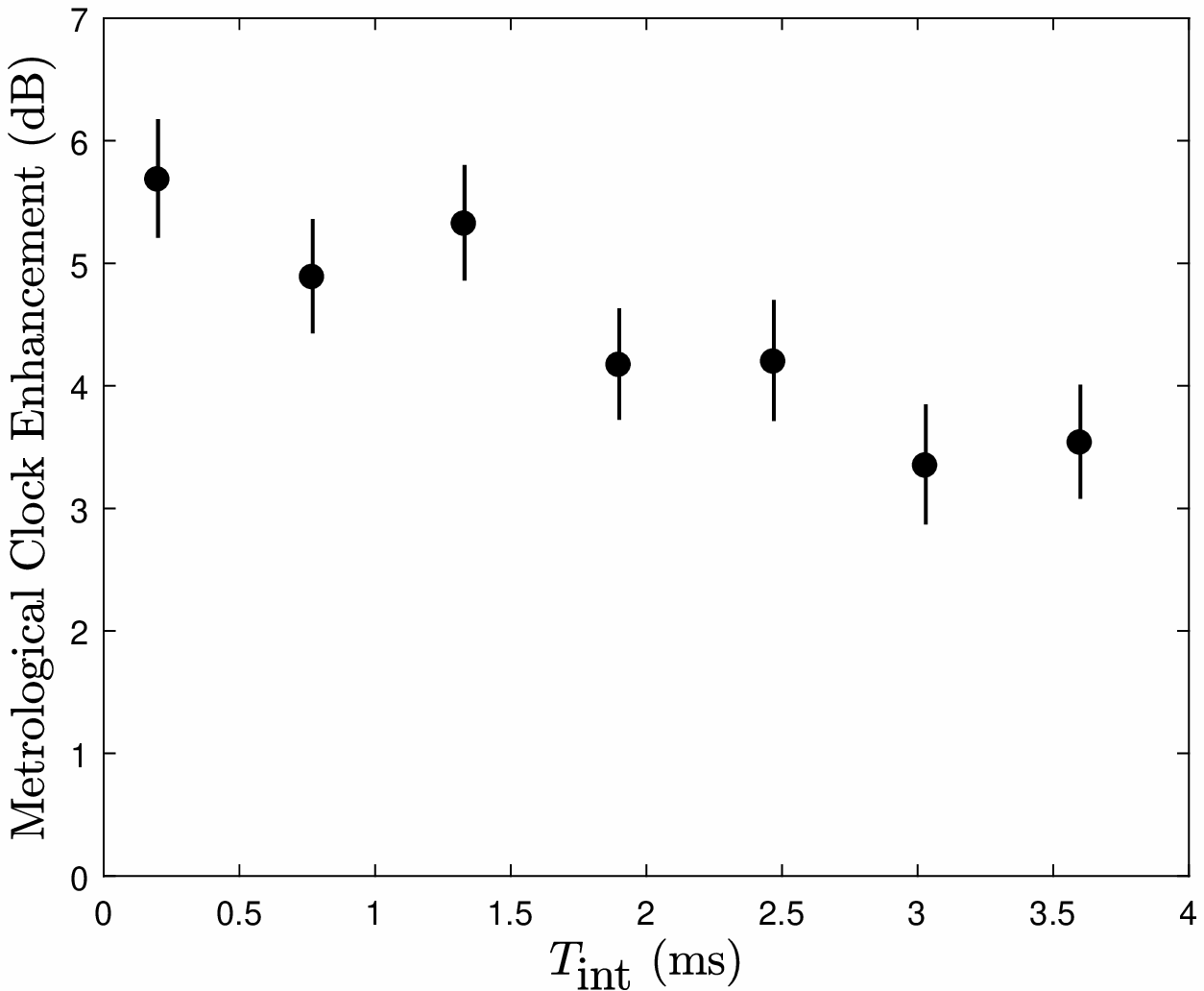}
\caption{Single-shot fractional stability ($\tau=T_c=1$ s) for several Ramsey times $T_{\text{int}}$ in a clock sequence with 240\,000 atoms. Error bars represent 68\% confidence intervals}
\label{fig:clock_enhancement}
\end{figure}

Metrologically-relevant squeezing can be calculated with respect to a shot-noise-limited frequency fractional stability,
\begin{equation}
    \sigma^{2(css)}_y(\tau) \coloneqq \text{var}(\bar{\phi}_\tau)=\frac{1}{\omega_0^2T_{int}^2}\frac{1}{N(\tau/T_c)},
\end{equation}
where $\bar{\phi}_\tau$ is the average of $\phi(t)$ over a time interval $\tau$. Full contrast is assumed. The expression takes such variance form due to the uncorrelated nature of the shot noise. Single-shot metrological improvement of the clock with reference to the QPN limit for a given $T_\text{int}$ is given as $\sigma_y(\tau=T_c)/\sigma_y^{(css)}(\tau=T_c)$ (Fig.~\ref{fig:clock_enhancement}). 

\section{Anti-squeezing}
Due to anti-squeezing, the final projection to the Jz axis of a state with non-zero $\theta$ creates a banana-like shape of the Wigner distribution on the Bloch sphere. This effect effectively increases the resolvable variance of the state as~\cite{Braverman_2018}
\begin{equation}
(\Delta\theta)^2 = \frac{\xi^2}{N} + \frac{(\Gamma^2-\Gamma^{-2})^2}{2N^2}\tan^2(\theta)+\Delta X_0,
\end{equation}
where $\xi^2 = -14$ dB is the amount of state squeezing, $\Gamma^2=37$ dB the amount of state anti-squeezing, and $\Delta X_0$ is the total camera noise at $\theta = 0$. Probe power dictates the amount of squeezing and anti-squeezing produced. Using less power will increase the maximum $\theta$ but at the cost of squeezing. We note that this equation describes our technical-limited results well. A good fit to the $\Delta\theta$ vs $\theta$ plot in the main document was obtained by leaving $\Gamma$, a free parameter. The calculated $\Gamma$ is consistent with a direct camera measurement in which the squeezed state was rotated $\pi/2$ about its center before detection.

\end{document}